\RequirePackage{fix-cm}
\documentclass[twocolumn]{svjour3}

\usepackage[utf8x]{inputenc}
\usepackage{color}
\usepackage{cite}
\usepackage{soul}
\usepackage{multicol}
\usepackage{multirow}
\usepackage{graphicx}
\usepackage{url}
\usepackage{amsmath}
\usepackage{flushend}
\usepackage[normalem]{ulem}
\usepackage[ruled,linesnumbered]{algorithm2e}
\usepackage{amssymb}
\usepackage{verbatim}
\usepackage{fancyvrb}

\PrerenderUnicode{å}

\journalname{International Journal on Software Tools for Technology Transfer}
\begin{document}

\definecolor{darkgreen}{RGB}{0,100,0}
\definecolor{orange}{RGB}{252,70,0}
\definecolor{grey}{RGB}{80,80,80}
\definecolor{navy}{RGB}{0,0,128}
\newif\ifcomments

\commentstrue

\newcommand{\commentColor}[3]{\ifcomments
~\textcolor{#3}{[\textbf{#1:} #2]}
\fi}
\newcommand{\commentIB}[1]{\commentColor{IB}{#1}{darkgreen}}
\newcommand{\commentVV}[1]{\commentColor{VV}{#1}{orange}}

\newcommand{\mycol}[2]{{\leavevmode\color{#1}#2}}
\newcommand{\temp}[1]{\operatorname{\mathbf{#1}}}
\newcommand{\q}{\textcolor{red}{\textbf{?}}}
\newcommand{\corr}[1]{\textcolor{navy}{#1}}
\newcommand{\x}[1]{\corr{#1}}

\newcommand{\myskip}{\noalign{\vspace{0.04cm}}}

\newcommand{\todo}[1]{\vspace{0.1cm}{\setlength{\parindent}{0pt}\framebox[3.31in][l]{\parbox{\dimexpr\linewidth-2\fboxsep-2\fboxrule}{\textcolor{red}{\textbf{TODO:} #1}}}}\vspace{0.1cm}}

\title{Combining closed-loop test generation and execution by means of model checking}

\author{Igor Buzhinsky$^{1, 2}$ \and Valeriy Vyatkin$^{1, 2, 3}$}
\authorrunning{Igor Buzhinsky, Valeriy Vyatkin}

\institute{
This work has been funded by the Government of the Russian Federation (Grant 08-08).
\vspace{0.25cm}\hrule width240pt\vspace{0.2cm}
Igor Buzhinsky (corresponding author)\\
igor.buzhinskii@aalto.fi\\
~\\
Valeriy Vyatkin\\
vyatkin@ieee.org\\
~\\
$^1$\hspace{0.15cm}Department of Electrical Engineering and Automation, Aalto University, Espoo, Finland\\
$^2$\hspace{0.15cm}Computer Technologies Laboratory, ITMO University, St. Petersburg, Russia\\
$^3$\hspace{0.15cm}Department of Computer Science, Electrical and Space Engineering, Lule{\aa} University of Technology, Sweden\\
}
\date{Received: date / Accepted: date}

\maketitle

\SetEndCharOfAlgoLine{}
\setcounter{footnote}{0}

\begin{abstract}
Model checking is an established technique to formally verify automation systems which are required to be trusted.
However, for sufficiently complex systems model checking becomes computationally infeasible.
On the other hand, testing, which offers less reliability, often does not present a serious computational challenge.
Searching for synergies between these two approaches, this paper proposes a framework to ensure reliability of industrial automation systems by means of hybrid use of model checking and testing.
This framework represents a way to achieve a trade-off between verification reliability and computational complexity which has not yet been explored in other approaches.
Instead of undergoing usual model checking, system requirements are checked only on particular system behaviors which represent a test suite achieving coverage for both the system and the requirements.
Then, all stages of the framework support the case of a closed-loop model, where not only the controller, but also the plant is modeled.
\end{abstract}

\section{Introduction}
\label{sec:introduction}

Ensuring reliability is a major issue related to mission-critical automation systems.
Speaking of functional requirements, two approaches to address this issue are known.
In \emph{simulation-aided testing}, the automation system is exercised on a finite number of behavior scenarios which are obtained according to some methodology, such as model-based testing (MBT)~\cite{broy2005model, rosch2015review}.
In \emph{formal verification by model checking}~\cite{clarke1999model}, behaviors of the formal model of the automation system are exhaustively explored.
These two approaches have strengths and weaknesses.
Simulation-aided testing is limited to particular behavior scenarios, but it is fast and does not require specific knowledge other than the domain one.
Thus, it is largely adopted in industry.
Model checking, on the other hand, is more reliable, but much more resource-intensive and requires expertise in formal methods.

Failures of mission-critical automation systems are not tolerated.
Thus, their functional correctness must be assured with formal methods.
Nowadays, they are used in certain cases~\cite{frey2000formal, ovatman2016overview, pakonen2017practical}, but far not ubiquitously.
Speaking of model checking, its application is especially complicated for large systems.
Modeling and verifying them straightforwardly would likely result in infeasible resource (CPU time and RAM) consumption.
Various solutions of this computational complexity problem are known, ranging from naive ones, like omission of certain aspects of the system or applying harsh discretization to real values, to more sophisticated abstraction techniques~\cite{kesten2000control, gourcuff2008improving} and compositional verification approaches~\cite{tripakis2016compositionality}.
Another solution, known as bounded model checking~(BMC)~\cite{biere2003bounded}, limits the complexity of considered behaviors of the formal model.

Unfortunately, in practical cases, approaches which preserve the thoroughness of model checking are often not sufficient, and one needs to sacrifice precision of either the model (e.g., omit some of its aspects) or the model checking algorithm (e.g., apply BMC) to make verification feasible.
On the other hand, applying several ``unsafe'' approaches wherein different aspects of model checking are affected may to some extent compensate the related decrease of reliability.
This motivates the development of new approaches of finding a trade-off between resource consumption and model checking reliability.

One area wherein such an approach does not yet exist is the combined use of testing and model checking.
Previously, multiple approaches of test generation by means of model checking were proposed~\cite{fraser2009testing, fraser2009issues, duarte2011integrating}.
Later, the process of test execution was represented as a model checking problem~\cite{buzhinsky2015formal, buzhinsky2017testing}.
Coupling these approaches in a single hybrid method is the topic of this paper.
By viewing all stages of automation system testing as model checking, the initial problem of model checking is transformed to the one of testing with a related decrease of reliability, but also with a significant performance improvement.
At the same time, the resultant approach is different from conventional testing by the use of temporal specification during both test case generation and execution.

More specifically, this paper proposes a framework to ensure reliability of industrial automation systems by hybrid use of testing and model checking.
The framework considers the general situation of a closed-loop system~\cite{preusse2013technologies}, where both the controller and the controlled plant are represented as formal models.
Extraction of nondeterminism of the plant model into designated input variables allows convenient representation and generation of test cases, which are then used to check temporal logic requirements specified for the closed-loop system.
Test cases are unwound into infinite paths, and thus the semantics of temporal logics is preserved.
On the other hand, test cases are selected according to coverage criteria for both the state space of the model and the subformulas of temporal requirements.
The framework has been implemented in a software tool which which is written in Java and is available online.\footnote{\url{https://github.com/igor-buzhinsky/formal_testing_in_closed_loop}}

The rest of the paper is structured as follows.
Section~\ref{sec:preliminaries} introduces used concepts and terms.
Section~\ref{sec:framework} describes the proposed framework.
Then, in Sections~\ref{sec:case_study}, the framework is applied on case studies and compared with conventional model checking.
In Section~\ref{sec:comparison}, it is compared with other verification approaches.
Section~\ref{sec:discussion} concludes the paper.

\section{Preliminaries}
\label{sec:preliminaries}

\subsection{Model checking of automation systems}

Model checking~\cite{clarke1999model} is a formal technique of state space analysis of a formal model.
Assume that the state of the system is time-dependent (the simplest case is logical, discrete time).
Then, if the formal model specifies valid initial states of the system and rules of state transition, it becomes possible to verify properties involving the state of the system in different time instants by performing exhaustive state space exploration.
Such properties are called \emph{temporal properties}.

In \emph{linear temporal logic (LTL)}, several \emph{temporal operators} allow formulating properties over system behaviors represented as infinite sequences of states.
Among temporal operators are $\temp{G}$ (always), $\temp{F}$ (eventually in the future) and $\temp{X}$ (in the next state).
For example, LTL property $\temp{X} \temp{G} (x \to \temp{F} y)$ specifies that starting from the second time instant, $x$ always implies $y$ in one of the future states (or in the same state where $x$ becomes true).
The \emph{problem of LTL model checking} is to determine whether the given LTL property is satisfied for all infinite behaviors of the given model, and, if not, provide a \emph{counterexample}---a behavior which violates this property.
Then, \emph{computation tree logic} (CTL) examines trees of possible system executions rather than separate behaviors.
Nevertheless, some temporal properties can be equivalently expressed in both CTL and LTL.

From now on, the system whose correctness must be ensured will be referred to as \emph{system under verification (SUV)}.
Model checking of industrial automation systems~\cite{frey2000formal, vyatkin2001formal, preuse2012closed, ovatman2016overview} is distinguished by the separation of the model of the SUV into the \emph{controller model} and the \emph{plant model}.
Often, only the former is prepared, resulting in \emph{open-loop model checking}.
However, modeling the plant (i.e., the devices with which the controller works and the corresponding physical processes) and, optionally, environmental conditions and intervention of human operators, allows considering the feedback between the plant and the controller.
The type of model checking, which enables controller verification in more natural conditions by considering the plant model, is known as \emph{closed-loop model checking}.
Many safety-critical properties in automation systems cannot be described in terms of the controller's input and output variables, but require reference to plant parameters, therefore closed-loop model checking is a more expressive and comprehensive method.

\subsection{Model checking algorithms and tools}

In \emph{explicit-state model checking}, the state space of the model is represented explicitly, as a \emph{state graph}, or, more precisely, as a Kripke structure~\cite{clarke1999model}.
Consequently, if the state space of the model is large, its processing becomes impossible due to excessive CPU time and, sometimes, RAM requirements.
This phenomenon is known as \emph{state space explosion}.
SPIN~\cite{holzmann1997model} is a well-known explicit-state model checker used primarily for verifying multi-process systems.
Its formal language is called Promela.
Promela is similar to an imperative programming language but allows nondeterministic statements.

\emph{Symbolic model checking}~\cite{burch1992symbolic} was suggested as a means of overcoming the state space explosion issue.
Its key idea is to process states implicitly, by operating with Boolean formulas.
One of the most popular symbolic model checkers is NuSMV~\cite{cimatti2000nusmv}.
Unlike Promela, NuSMV models are specified as systems of constraints over state variables of the SUV.


\subsection{Reducing model checking complexity}

A common way to mitigate model checking complexity is \emph{abstraction}, a simplification of the SUV which preserves the results of model checking or alters them in a tolerable way.
Multiple abstraction approaches have been proposed.
For example, the work~\cite{kesten2000control} proposed two kinds of abstraction: control abstraction allows handling of systems with unbounded structure, and data abstraction maps variables over infinite domains to finite ones.
In particular, control abstraction requires the SUV to be \emph{modular}.
According to~\cite{tripakis2016compositionality}, another use of modularity is to avoid repeated verification of the entire SUV when one of its components is refined.

In~\cite{gourcuff2008improving}, abstractions are proposed specifically for PLCs (programmable logic controllers), which are distinguished by cyclic operation, wherein a single cycle involves reading inputs, executing the PLC program, and updating outputs.
One of the abstractions skips intermediate states of the PLC cycle, leaving only the final values of output variables.
The other one reduces the number of variables which describe the state of the PLC.
These abstractions are applicable when the verified properties deal with extrinsic PLC behavior, which is composed of input/output values in the end of each cycle.

The alternative solution to handling the complexity issue is to change the way how temporal properties are checked instead of modifying the model of the SUV.
\emph{Partial order reduction} and \emph{cone of influence reduction} are implemented in model checkers SPIN~\cite{holzmann1997model} and NuSMV~\cite{cimatti2000nusmv} respectively.
\emph{Bounded model checking} (BMC)~\cite{biere2003bounded} is a technique of symbolic model checking where the LTL model checking problem is reduced to finding a satisfying assignment of a propositional formula by executing a satisfiability (SAT) solver.
The reduction is done with the following limitation: only counterexamples which can be represented with a state sequence of the given length $k$ are considered (note that counterexamples still can be infinite since this sequence may contain a cycle).
This means that an LTL property may be falsely reported to be satisfied if the given $k$ is insufficient.
NuSMV is able to perform BMC by incrementally increasing $k$ up to the given limit.


\subsection{Testing with model checkers}
\label{sec:testing_with_mc}

Test case generation by model checkers is a widely explored topic, which is thoroughly reviewed in~\cite{fraser2009testing}.
The main idea of such test case generation is to (1) formulate \emph{test purposes} which describe the desired properties of test cases; (2) formulate temporal properties expressing that test purposes are never reached---so-called \emph{never claims}; and (3) model-check never claims.
The result of model checking a never claim is either a counterexample, which is interpreted as a test case (it shows that the never claim is violated, i.e., the test purpose is reached), or no counterexample, meaning that the corresponding test purpose is unreachable.

One of the ways to formulate test purposes is to base them on \emph{coverage}~\cite{broy2005model, fraser2009testing}, a popular property which is often required for test suites in industry.
Multiple coverage criteria are known, including structural ones (e.g., whether a particular instruction or line of code is executed, or a particular condition in a conditional statement is realized) and data-based ones (e.g., certain predicates over the state space of the system are satisfied).
Another idea is to make generated test cases distinguish the correct program and its mutants, which are obtained by making minor syntactic changes in the program.
Coverage criteria can be also based on data flow~\cite{su2015combining} or temporal specification~\cite{tan2004specification, zeng2012test}.
Much more approaches are reviewed in~\cite{fraser2009testing}.
Unfortunately, as these approaches become more complex and require generation of more test cases, the time to generate the entire test suite also grows.
As a partial remedy, BMC can be used to generate test cases faster~\cite{heimdahl2003auto, angeletti2010using}.


The works~\cite{buzhinsky2017testing, buzhinsky2015formal} apply model checkers to deal with test case execution rather than generation.
In these works, which will be discussed in more detail in Section~\ref{sec:comparison}, the testing problem is limited to automation systems.

However, by now, there is no approach which is based on model checking and encompasses all the stages of testing.
Test case generation by model checking neither considers the \emph{test oracle} problem~\cite{barr2015oracle}, that is, distinguishing correct results of test executions from incorrect ones, nor the question of test case execution.
On the other hand, the works~\cite{buzhinsky2017testing, buzhinsky2015formal} do not propose integrated approaches of test generation and instead assume that test cases and their oracles are obtained by other means (e.g., using MBT).
Coupling these two research direction into a single framework is the intended contribution of the present work.

\section{Proposed framework}
\label{sec:framework}


\subsection{Plant, controller models and nondeterminism}
\label{sec:var_types}

A \emph{Mealy machine} is a tuple $(S, s_0, I, O, \delta, \lambda)$.
Here, $S$ in a finite set of \emph{states}, among which $s_0 \in S$ is the \emph{initial state}.
Then, $I$ is a set of \emph{input variables}, each of which has a finite set of possible values (e.g., Booleans or integers), and $O$ is a set of \emph{output variables}.
Finally, $\delta : S \times v(I) \to S$ is the \emph{transition function} and $\lambda : S \times v(I) \to v(O)$ is the \emph{output function}, where $v(I)$ and $v(O)$ are sets of input and output variable value combinations respectively.

Both the plant and the controller models are viewed as Mealy machines which pass their output variables as input variables to each other.
Technically, such state machines can be specified in modeling languages such as NuSMV or SPIN.
From now on, the terms \emph{input variables} and \emph{output variables} will be referred to the ones of the controller.
The interaction of the plant and the controller models is cycle-based: first the plant model makes its turn, and then the controller model does.

However, if both models are represented as deterministic state machines, this means that the closed-loop model of the SUV has only one behavior.
To resolve this problem, nondeterminism is modeled as arbitrary selection of values of special variables, from now on referred to as \emph{nondeterministic variables}.
In the most natural cases, these variables correspond to inputs from human operators or random behavior of the plant.
They are passed as input to the plant model along with the output variables of the controller.
This approach of extracting nondeterminism into independent variables is useful throughout the framework.

\begin{figure*}[!t]
\centering
\includegraphics[width=6.5in]{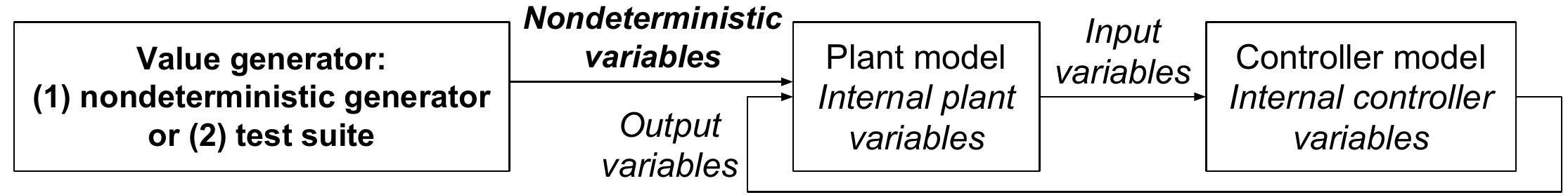}
\caption{Model interaction overview and variable kinds. The elements of the diagram which differ the framework from the traditional closed-loop verification methodology are shown in bold}
\label{fig:model_interaction}
\end{figure*}

\begin{figure*}[!t]
\centering
\includegraphics[width=6.5in]{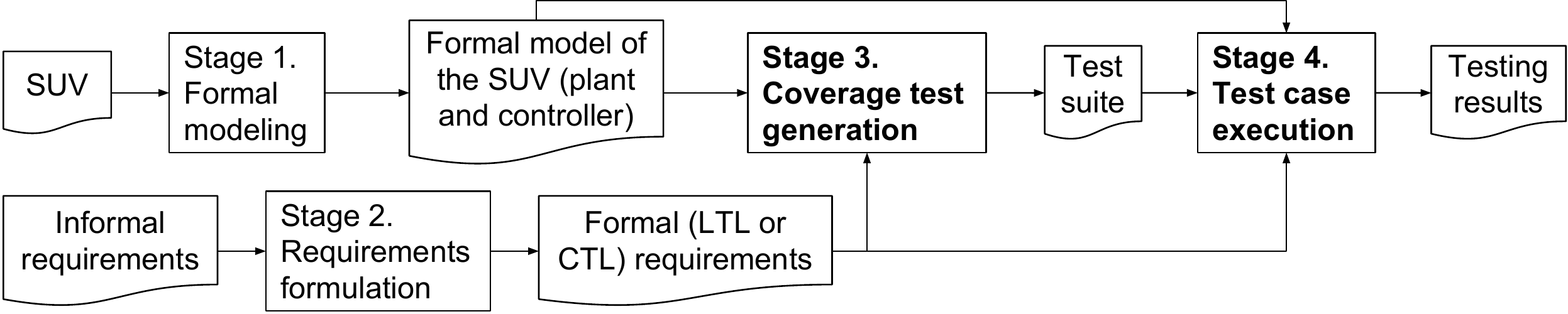}
\caption{Artifacts and stages of the framework. The stages which comprise the novelty of the framework are shown in bold}
\label{fig:stages}
\end{figure*}

The overview of considered models and their interaction by variables is shown in Fig.~\ref{fig:model_interaction}.
In addition to previously mentioned variable kinds, plant and controller models have their own \emph{internal variables}, which comprise their states.
Then, the values of nondeterministic variables may originate from two sources, or two kinds of \emph{value generators}: either (1) a purely nondeterministic generator (natural for ordinary model checking where the whole range of possible system behaviors is explored), or (2) a deterministic one (to enable execution of deterministic test cases which are encoded in these values).

Finally, although the framework is intended for closed-loop use, the plant model can be omitted to make it applicable in the open-loop scenario.
This can be done by specifying a dummy plant model which copies the values of nondeterministic variables to input ones, making these kinds of variables identic.
Thus, plant behaviors become arbitrary, which corresponds to the open-loop assumptions.

\subsection{Test suite model}
\label{sec:ts_model}

The most notable kind of a value generator is a test suite model.
Let $v_1, ..., v_k$ be the nondeterministic variables.
Assume that their values belong to finite sets $V(v_1), ..., V(v_k)$ respectively.
A \emph{test case} of length $\ell$ is a matrix $\{T_{i, j}~|~1 \le i \le k, 1 \le j \le \ell\}$, where $T_{i, j} \in V(v_i)$ specifies the value of $v_i$ on step $j$.
Thus, a test case specifies a concrete selection of nondeterministic variable values, which fixes plant behavior.
However, unlike conventional testing, model checking considers infinite behaviors of systems under examination.
While obtained as finite sequences (Section~\ref{sec:synthesis}), test cases will be executed (Section~\ref{sec:execution}) as infinite ones by looping the finite sequence from its beginning (such looping satisfies the need to have infinite paths to check temporal properties).
Encoding such looping behavior in model checkers NuSMV and SPIN (i.e., producing textually represented test case models which precisely represent test cases) is straightforward: the next step of the test case is calculated as $(j \text{ mod } \ell) + 1$.

In the definition of the test case above, we have neglected the test oracle problem.
The solution developed in this paper is to use temporal specifications for this purpose---the ones which would have been otherwise used for ordinary model checking.
For a fixed SUV, such oracles are independent from concrete test cases.

A \emph{test suite} is a set of test cases.
The corresponding test suite model (i.e., exact test suite representation in a formal language such as NuSMV or SPIN) works as follows, behaving as a value generator.
Before the first combination of values is generated, the test case to be executed is selected nondeterministically (technically, this is done by assigning an index to each test case and nondeterministically selecting this index).
Then, the values of the chosen test case are generated on each step, being looped over as described above.
As a result, recalling that every other model in the closed-loop system is deterministic, the test suite model causes the system's state graph to be formed of a number of independent infinite paths.
Model checking algorithms, nevertheless, can process such graphs in finite time since each infinite path has a lasso shape~\cite{clarke1999model}, i.e., a finite prefix followed by a loop.

\subsection{Stages of the framework}
\label{sec:stages}
The proposed framework comprises four stages:
\begin{enumerate}
\item \textbf{Formal modeling.}
As a prerequisite of applying the framework, one needs to prepare the formal models of the plant and the controller.
The controller model may be generated automatically based on the original controller code (e.g., as in~\cite{darvas2017plc}), and the plant model can be not only created manually, but also constructed based on system behavior traces (e.g., as in~\cite{maier2011anomaly, buzhinsky2017automatic}).
As explained below, these models must be available in both a symbolic verifier (e.g., NuSMV) and an explicit-state verifier (e.g., SPIN).
\item \textbf{Requirements formulation.}
Functional requirements to the SUV are formulated in temporal logics, such as LTL or CTL.
This process is identical to the one commonly done in model checking of automation systems and thus is not discussed further.
These temporal requirements will be used as test oracles for test cases generated on the next stage.
Note that this stage requires expertise in formal methods.
\item \textbf{Coverage test generation.}
The value generator of type~1 (according to Fig.~\ref{fig:model_interaction}) produces the values of all nondeterministic variables arbitrarily and independently.
During a symbolic model checking run, this is needed to find sequences of values which achieve coverage of the SUV's state space and subformulas of temporal requirements.
These sequences form the test suite.
\item \textbf{Test case execution.}
Test cases found on the previous stage are executed by using the value generator of type~2 (according to Fig.~\ref{fig:model_interaction}) during an explicit-state model checking run.
As a result, some of them may be reported as counterexamples to the requirements formulated on stage~2.
\end{enumerate}

Stages~3 and~4, which represent the novelty of the framework, are examined in detail below.
All the stages are also presented in Fig.~\ref{fig:stages} together with their inputs and outputs.

\subsection{Coverage test generation}
\label{sec:synthesis}

In Section~\ref{sec:testing_with_mc}, the general idea of test case generation with model checking was mentioned together with several approaches of selecting test purposes.
The coverage test generation stage of the proposed framework adapts these ideas to the considered problem.
Since both a large number of test cases to be generated and the complexity of test case generation criteria influence the time required to generate the test suite, a reasonable limitation of both factors is needed: otherwise, the proposed approach would fail to compete with other model checking complexity reduction approaches in performance.
On the other hand, since the considered SUVs are closed-loop, not only the coverage of the controller model can be assured, but also the one of the plant model.
Moreover, according to the principles of MBT, coverage of the specification is also helpful.
To sum up, the chosen test purposes need to cover all available artifacts (plant model, controller model, specification) while remaining lightweight.

The most straightforward test purposes capable of covering each of the artifacts are data-based.
Suppose that $p$ is a Boolean predicate of the state space of the SUV.
The coverage goal to reach $p$ can be formulated in LTL as $\temp{G} \neg p$, where $\temp{G}$ means ``always''.
Speaking of the coverage of the state space of the closed-loop model, the simplest way to select $p$ is according to \emph{state coverage}: $p = (v = v_0)$, where $v$ is a variable and $v_0 \in V(v)$.

The same principle can be applied to specification coverage.
In MBT, the formal model of requirements is usually represented as a state machine or a system of pre- and postconditions and is covered structurally.
In our case, structural coverage of temporal requirements can be interpreted as coverage of their subformulas.
Previously, an approach~\cite{tan2004specification} has been proposed to achieve subformula coverage of LTL formulas based on the idea of refuting formula mutations by generated test cases.
However, this approach is too computationally intensive for our purposes since it involves running a model checker multiple times to check formulas whose complexity is similar to the LTL specification.
Thus, the complexity of coverage test generation becomes comparable with the one of actual model checking, which we intend to avoid in our testing framework.
To achieve better performance, we use a simpler approach which only involves coverage of Boolean subformulas (i.e., the subformulas of the LTL formula which do not contain temporal operators): if $f$ is a Boolean subformula of one of temporal (either LTL or CTL) requirements of the SUV, then $p \in \{f, \neg f\}$ (to cover both true and false values of $f$).

The ideas given above are summarized in the test suite generation algorithm (Alg.~\ref{alg:synthesis}) and are explained in detail below.
For each variable $v$ in the system (of either of five variable kinds) and for each value $v_0$ in the set of $v$'s possible values $V(v)$, never claim $\temp{G} (v \ne v_0)$ is considered (lines~2--4).
Similarly, a never claim is considered for both values of each Boolean subformula of each temporal requirement (lines~5--7).
Then, a set of processed never claims is maintained.

For each never claim which is not yet processed, a model checking run is performed to verify this never claim (line~10).
We assume that plant and controller models follow the idea of interpretation abstraction~\cite{gourcuff2008improving}: a single step of the overall formal model corresponds to a single execution of both the plant and the controller models viewed as Mealy machines (for example, if the controller represents a PLC program, then all its internal assignments are atomic).
This abstraction is extremely beneficial for test execution and allows representing temporal properties more conveniently.
The interaction of elements shown in Fig.~\ref{fig:model_interaction} is implemented as follows: on each discrete time step, the nondeterministic generator produces a tuple of values of nondeterministic variables.
In symbolic verifier NuSMV, such a generator has an empty implementation since no restrictions are placed on these variables.
These values, together with previous-step output values of the controller (or default values on the first step) are passed to the plant model, which produces an input tuple for the controller.
The controller model transforms these inputs into outputs.

If the model checking run is performed using BMC, then it depends on the bound $k = \ell_{\max} - 1$.
This run either detects that the never claim is violated and returns a counterexample $c$ representing a test case reaching the coverage goal, or proves this claim (returns \emph{null}), which means that no test case limited by $\ell_{\max}$ exists which reaches this goal.
Regardless of the result of model checking, this never claim is marked as processed (line~11).
Then, if the test case has been found (line~12), it is added to the test suite (line~13).
It might also happen that the found test case covers some other unprocessed coverage goals---in this case such goals are not considered further (lines~14--18).
Function $\mathtt{covers}$ is implemented in a way which does not involve execution of $c$, and thus checking for coverage of other goals is faster than trying to generate a test case for each coverage goal.

\begin{algorithm}
 \SetKwData{processed}{processed}\SetKwData{neverClaims}{neverClaims}
 \SetKwFunction{bmcverif}{modelCheck}\SetKwFunction{covers}{covers}
 \KwData{model $S$ with a nondeterministic value generator, set $\mathcal{V}$ of all variables in $S$, maximum test case length $\ell_{\max}$, set $R$ of temporal requirements}
 \KwResult{test suite $\mathcal{T}$}
 $\neverClaims \leftarrow \varnothing$; $\processed \leftarrow \varnothing$; $\mathcal{T} \leftarrow \varnothing$;\\
 \ForEach{$v \in \mathcal{V}$, $v_0 \in V(v)$}{
     $\neverClaims \leftarrow \neverClaims \cup \{\temp{G} (v \ne v_0)\}$;
 }
 \ForEach{$r \in \mathcal{R}$, \upshape Boolean subformula $f$ of $r$}{
     $\neverClaims \leftarrow \neverClaims \cup \{\temp{G} f\} \cup \{\temp{G} \neg f\}$;
 }
 \ForEach{$p \in \neverClaims$}{
    \If{$p \notin$ \processed}{
        $c \leftarrow$ \bmcverif{S, p, $k = \ell_{\max} - 1$};\\
        $\processed \leftarrow \processed \cup \{p\}$;\\
        \If{$c \ne$ null}{
            $\mathcal{T} \leftarrow \mathcal{T} \cup \{c\}$;\\
            \ForEach{$p' \in \neverClaims \setminus \processed$}{
                \If{\covers{$c, p'$}}{
                    $\processed \leftarrow \processed \cup \{p'\}$;\\
                }
            }
        }
    }
 }
\caption{Test suite generation algorithm}
\label{alg:synthesis}
\end{algorithm}

Above, test cases have been mentioned to be generated by means of model checking.
One may execute explicit-state or symbolic BDD-based model checking algorithms, the advantage of which is the ability to generate counterexamples or prove their absence regardless of the bound on length $\ell_{\max}$.
Symbolic model checking is superior in this case as the one capable of handling large state spaces, while in explicit-state model checking the number of states in the model grows exponentially with the test case length required to achieve coverage.
Then, it was found that applying BDD-based exact symbolic model checking for test case generation is comparable with usual BDD-based symbolic model checking of the closed-loop system in terms of required time.
Thus, the goal of applying the proposed framework to reduce model checking time would not be reached.
Consequently, instead of BDD-based model checking, we apply BMC, which has also been reported to be efficient for test case generation in~\cite{heimdahl2003auto, angeletti2010using}.
In this case the bound $k = \ell_{\max} - 1$ becomes relevant.
In the tool implementing the framework, symbolic generation of test suites is implemented by utilizing NuSMV.

\subsection{Test suite execution}
\label{sec:execution}

When the test suite is generated, it remains to run the test cases against the requirements formulated on the second stage of the framework (Section~\ref{sec:stages}).
Similarly to test case generation and developing the ideas of~\cite{buzhinsky2015formal, buzhinsky2017testing}, this stage is also performed by means of model checking.

On each step of model execution, a new tuple of nondeterministic values is produced by the test suite model (as described in Section~\ref{sec:ts_model}), which is represented in a language of an explicit-state model checker.
This model, implementing a value generator of type~2 according to Fig.~\ref{fig:model_interaction}, exhibits nondeterminism only on the first step, when the test case is selected.
Then, the extraction of nondeterministic variables into a separate variable kind becomes useful: it makes model execution deterministic once their values are fixed and the test case is selected, and thus subsequent model checking is equivalent to conventional test case execution.

The rest of value passing and execution of plant and controller models is similar to the one in the test case generation stage.
However, the purpose of model checking is now different.
During test case generation, a path (specified by the values of nondeterministic variables) was searched which achieves a particular coverage goal.
In test case execution, the purpose is to check whether the requirements of the SUV are satisfied on test cases, which are viewed as infinite extensions of values sequences found on the previous stage.
Thus, if a counterexample is found by the model checker, it corresponds to one of the test cases.
If the model checker finds no counterexample, this means that the requirement is satisfied for the entire test suite.
The described procedure is repeated for each requirement of the SUV.

We stress the importance of performing test case execution in an explicit-state model checker (in contrast to test case generation), since this makes verification significantly faster~\cite{buzhinsky2017testing}.
This is explained by determinism of test cases, which prevents state space explosion: the model checking algorithm examines only a small number of lasso-shaped paths.
However, the need to perform explicit-state model checking in addition to symbolic one implies the need to model the SUV in the languages of both a symbolic verifier (such as NuSMV) and an explicit-state verifier (such as SPIN).
In the tool implementing the framework, test case execution is implemented by running SPIN on test cases represented in Promela.

\section{Case studies}
\label{sec:case_study}

\begin{figure}[!t]
\centering
\includegraphics[width=2.8in]{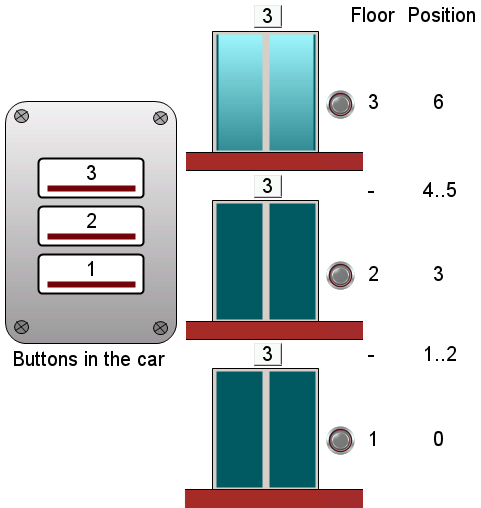}
\caption{Visualization of the elevator simulation model (the car is at floor 3)}
\label{fig:elevator}
\end{figure}

We evaluate the proposed framework on two case studies.
The first case study used in this paper comprises an elevator control system.
It is based on a Structured Text (ST) PLC application implemented in CODESYS\footnote{\url{https://www.codesys.com/}} together with the elevator model and its visualization (Fig.~\ref{fig:elevator}).
The elevator is designed for a three-story building.
Every floor of the building has a button to call the elevator; three buttons (for each floor) are also located in the elevator's car.
When a button is pressed, it remains on (i.e., cannot be pressed again) until the elevator arrives at the designated floor and its doors are opened.
The controller implements this very behavior; however, when multiple floors are requested infinitely, the implementation of the controller prohibits some floors from being reached (for example, when the car is at floor 2 and floors 1 and 3 are called, the car will always go up).
When the doors open, they remain open for some time before closing.

\begin{figure}[!t]
\centering
\includegraphics[width=3.25in]{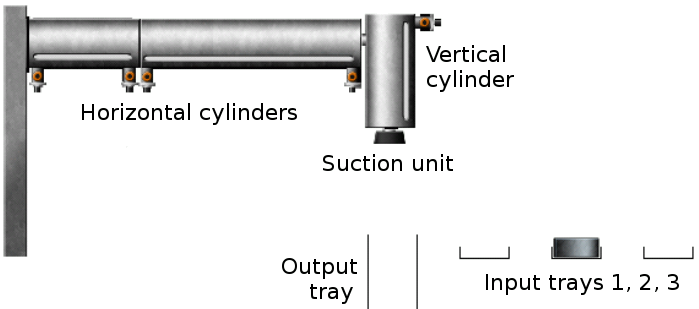}
\caption{Visualization of the pick-and-place simulation model (the manipulator is in its initial position, and a workpiece is present on input tray 2)}
\label{fig:pnp}
\end{figure}

The second case study is a control system for a pick-and-place manipulator (PnP).
A similar system has been previously used in~\cite{patil2015counterexample}.
The visualization of the PnP system in nxtSTUDIO\footnote{\url{https://www.nxtcontrol.com/en/engineering/}} is shown in Fig.~\ref{fig:pnp}.
The plant comprises three input trays, on which workpieces can be placed from outside of the system, an output tray, which is the destination of all input workpieces, and the manipulator itself.
The manipulator comprises two horizontal pneumatic cylinders and one vertical cylinder, all of which are capable of extension and retraction, and a suction unit.
The implementation of the controller from our case study, which has been prepared in ST, works as follows: if there is a workpiece on some input tray, the workpiece from the tray with the minimum number is taken (only the trays with workpieces are counted), and then placed on the output tray.

\subsection{Formal models}

NuSMV and Promela models for the case studies were prepared both manually and by using automatic model code generation.
They employ discrete elevator car and cylinder positions and logical time.
For generated models, this discretization was implemented by simplifying ST source code.
Since Promela is an imperative language, like ST, automatic generation of Promela models is straightforward and thus differences between manual and automatically generated Promela models were insignificant.
In contrast, for NuSMV, which is a declarative language, such a conversion is more complex.
Although some NuSMV model generation techniques were proposed~\cite{gourcuff2006efficient, darvas2017plc}, they are not publicly available, and hence we implemented our own converter to handle the subset of ST that we use.
This converter is less efficient than the one in~\cite{darvas2017plc}, which leads to a practical outcome: the ability to evaluate the proposed framework on NuSMV models which are hard for NuSMV to process.

To evaluate the proposed framework on problems of varying complexity, the models were generalized to support arbitrary complexity value $n$, which corresponds to the number of floors (elevator case study), or the number of horizontal cylinders (PnP case study).
This was done by producing models by scripts which take $n$ as a parameter.
In the case of automatically generated models, the simplified ST code was produced by such a script, and only then the models were generated.

Below, we consider elevator models with $3 \le n \le 15$ floors.
In each model, the range of elevator car positions is $0..3(n - 1)$, where position $3(i - 1)$ corresponds to floor $i$ and other positions are intermediate.
Then, one logical time step corresponds to one second, and the elevator car moves with the speed of one position per time step.

PnP models are only considered for $2 \le n \le 5$ due to their exponential complexity: a model with $n$ horizontal cylinders is able to reach $2^n$ different trays, $2^n - 1$ of which are input ones.
This requires supporting a longer range of positions for each new horizontal cylinder: for the model with $n$ cylinders, the number of possible positions of the longest cylinder is $2^{n - 1} + 1$.
The vertical cylinder always has only three positions (retracted, intermediate, extended).
Similarly to elevator models, each cylinder moves with the speed of one position per time step.

The complexity of models (for elevator models, only for several values of $n$) is given in Table~\ref{tab:complexity}.
State space sizes, which are equal for manual and generated models, were computed with NuSMV.
Section~\ref{sec:comp_with_mc} will also clarify the complexity of models in symbolic model checking.



\begin{table*}
\centering
\caption{Complexity of case studies}
\renewcommand{\arraystretch}{1.3}
\label{tab:complexity}
\begin{tabular}{rrrrrrrrrrrrrrrr}
\hline\myskip
& \multirow{2}{*}{$n$} & \hspace{-0.3cm} & \multicolumn{4}{c}{Number of variables} & \hspace{-0.3cm} & \multicolumn{2}{c}{LOC$^\text{c}$, manual models} & \hspace{-0.3cm} & \multicolumn{2}{c}{LOC$^\text{c}$, generated models} & \hspace{-0.3cm} & State space\\
& & \hspace{-0.3cm} & N/d$^\text{a}$ & Input & Output & Internal$^\text{b}$ & \hspace{-0.3cm} & NuSMV & Promela$^\text{d}$ & \hspace{-0.3cm} & NuSMV & Promela & \hspace{-0.3cm} & \multicolumn{1}{c}{size}\\
\cline{1-2}\cline{4-7}\cline{9-10}\cline{12-13}\cline{15-15}\myskip
\multirow{5}{*}{\rotatebox[origin=c]{90}{Elevator}} &
   3 & \hspace{-0.3cm} &  6 & 15 &  5 &  8 & \hspace{-0.3cm} &  58 &  76 & \hspace{-0.3cm} &  154 &  209 & \hspace{-0.3cm} & $1.2 \cdot 10^4\hspace{0.145cm}$\\
&  6 & \hspace{-0.3cm} & 12 & 30 &  8 & 11 & \hspace{-0.3cm} &  97 &  76 & \hspace{-0.3cm} &  319 &  470 & \hspace{-0.3cm} & $1.8 \cdot 10^7\hspace{0.145cm}$\\
&  9 & \hspace{-0.3cm} & 18 & 45 & 11 & 14 & \hspace{-0.3cm} & 136 &  76 & \hspace{-0.3cm} &  538 &  839 & \hspace{-0.3cm} & $2.1 \cdot 10^{10}$\\
& 12 & \hspace{-0.3cm} & 24 & 60 & 14 & 17 & \hspace{-0.3cm} & 175 &  76 & \hspace{-0.3cm} &  811 & 1316 & \hspace{-0.3cm} & $2.0 \cdot 10^{13}$\\
& 15 & \hspace{-0.3cm} & 30 & 75 & 17 & 20 & \hspace{-0.3cm} & 214 &  76 & \hspace{-0.3cm} & 1138 & 1901 & \hspace{-0.3cm} & $1.8 \cdot 10^{16}$\\
\cline{1-2}\cline{4-7}\cline{9-10}\cline{12-13}\cline{15-15}\myskip
\multirow{4}{*}{\rotatebox[origin=c]{90}{PnP}} &
  2 & \hspace{-0.3cm} &  3 & 10 &  4 &  9 & \hspace{-0.3cm} &  71 &  95 & \hspace{-0.3cm} & 116 & 123 & \hspace{-0.3cm} & $5.7 \cdot 10^2\hspace{0.145cm}$ \\
& 3 & \hspace{-0.3cm} &  7 & 16 &  5 & 10 & \hspace{-0.3cm} &  91 &  95 & \hspace{-0.3cm} & 151 & 156 & \hspace{-0.3cm} & $1.1 \cdot 10^5\hspace{0.145cm}$\\
& 4 & \hspace{-0.3cm} & 15 & 26 &  6 & 11 & \hspace{-0.3cm} & 119 &  95 & \hspace{-0.3cm} & 206 & 205 & \hspace{-0.3cm} & $2.6 \cdot 10^9\hspace{0.145cm}$\\
& 5 & \hspace{-0.3cm} & 29 & 42 &  7 & 12 & \hspace{-0.3cm} & 163 &  95 & \hspace{-0.3cm} & 301 & 286 & \hspace{-0.3cm} & ---$^\text{e}$\\
\hline
\multicolumn{15}{l}{\scriptsize $^{\text{a}}$Nondeterministic. $^{\text{b}}$Sum of the numbers of plant and controller internal variables; the value is shown only for generated models.}\\
\multicolumn{15}{l}{\scriptsize $^{\text{c}}$Lines of code, sum for plant and controller models. $^{\text{d}}$Constant since loops over all floors, cylinders, or workpieces were used.}\\
\multicolumn{15}{l}{\scriptsize $^{\text{e}}$Unknown since time limit (24 hours) was exceeded.}
\end{tabular}
\end{table*}

\subsection{Temporal requirements}
\label{sec:requirements}

\subsubsection{Elevator case study}

For each elevator model, a set of the following $4n$ temporal requirements was generated in both LTL and CTL (floor index $i$ ranges from 1 to $n$):
\begin{enumerate}
\item ERT$_1^i$: always, if the car is between floors, the doors of floor $i$ shall be closed;
\item ERT$_2^i$: always, if either of the buttons of floor $i$ is pressed, either the car shall eventually arrive at floor $i$ and the doors of floor $i$ shall open, or one of the buttons of other floors is pressed;
\item ERT$_3^i$: always, if the doors of floor $i$ become open, they shall remain open for two additional time steps and then start closing;
\item ERT$_4^i$: always, if the doors of floor $i$ start closing and either of the buttons of floor $i$ is pressed, the doors shall reopen in two times steps.
\end{enumerate}

To properly evaluate the framework, requirements which were supposed to be violated were also prepared (differences from the previous requirements are written in italic):
\begin{enumerate}
\item ERF$_1^i$: always, if the car is between floors, the doors of floor $i$ shall be \emph{open};
\item ERF$_2^i$: always, if either of the buttons of floor $i$ is pressed, the car shall eventually arrive at floor $i$ and the doors of floor $i$ shall open \emph{(regardless of other buttons being possibly pressed)};
\item ERF$_3^i$: always, if the doors of floor $i$ become open, they shall remain open for \emph{one additional time step} and then start closing;
\item ERF$_4^i$: always, if the doors of floor $i$ start closing and either of the buttons of floor $i$ is pressed, the doors shall reopen in \emph{one time step}.
\end{enumerate}

\subsubsection{Pick-and-place manipulator case study}

For each PnP model, the following requirements were specified for each input tray (input tray index $j$ ranges from 1 to $2^n - 1$):
\begin{enumerate}
\item PRT$_1^j$: always, if a workpiece is on input tray $j$, it will be eventually taken, then eventually put on the output tray, and then the manipulator will eventually return to its initial position;
\item PRT$_2^j$: always, if a workpiece is on input tray $j$, this tray will eventually be empty, unless a new workpiece is added to this tray;
\item PRT$_3^j$: the cylinders of the manipulator are not positioned to take a workpiece from input tray $j$ until a workpiece appears on it, or are never positioned like that if it never appears.
\end{enumerate}

Then, these requirements were modified to make them violated:
\begin{enumerate}
\item PRF$_1^j$: always, if a workpiece is on input tray $j$, it will be eventually taken, then eventually put on the output tray, and then the manipulator will eventually \emph{retract the vertical cylinder and extend some horizontal cylinder};
\item PRF$_2^j$: always, if a workpiece is on input tray $j$, this tray will eventually be empty \emph{(regardless of whether this workpiece is further added again)};
\item PRF$_3^j$, $1 \le j < 2^n - 1$: the \emph{horizontal} cylinders of the manipulator are not positioned to take a workpiece from input tray $j$ until a workpiece appears on it, or are never positioned like that if it never appears \emph{(this requirement is false since the manipulator may be attempting to take a workpiece with a larger index)}.
\end{enumerate}

Note that the number of requirements is exponential of $n$.
For $n = 5$, this makes testing and verification sufficiently complex to compensate the relative simplicity of this PnP model compared to the elevator model with $n = 15$ (see Table~\ref{tab:complexity}).

\subsection{Example of a formal model}

The listing below provides an example of a Promela elevator model for $n = 3$ floors, which is suitable for test suite execution in SPIN.
It includes variable declarations (according to variable kinds described in Section~\ref{sec:var_types}), a test suite model with two test cases, each with two elements, a manually prepared plant model, and a stub of the controller model (the entire code is not shown due to its length).
The small size of the test suite was chosen only for the purpose of demonstration; the framework uses much larger case studies as explained in Section~\ref{sec:comp_with_mc}.
The variables of the test suite model comprise the index of the test case in the test suite (\texttt{\_test\_index}) and the current step of test case execution (\texttt{\_test\_step}).
Then, the main Promela process \texttt{init} is declared as an infinite \texttt{do} loop of atomic steps (intermediate computations within these steps are invisible in model checking).
Each step comprises test case selection (this is done only once per SUV run), setting nondeterministic variables according to the selected test case, incrementing test case step, and finally executing plant and controller models.
The LTL formula in end of the listing corresponds to ERT$_1^1$.
Note that array indexing in Promela is zero-based, and thus floors, test cases and their steps in the listing below are indexed from zero rather than one.

\newcommand\lparen[1]{(}
\newcommand\rparen[1]{)}
\begin{Verbatim}[commandchars=\\(),fontsize=\footnotesize]

// List of door states
\textbf(mtype) {d_closed, d_opening, d_open, d_closing}
// Input variables
\textbf(bool) on_floor[3] = 0;
\textbf(bool) door_closed[3] = 0, door_open[3] = 0;
\textbf(bool) button[3] = 0, call[3] = 0;
// Output variables
\textbf(bool) up = 0, down = 0, open[3] = 0;
// Nondeterministic variables
\textbf(bool) user_floor_button[3] = 0;
\textbf(bool) user_cabin_button[3] = 0;
// Plant internal variables
\textbf(int) elevator_pos = 0;
\textbf(mtype) door_state[3] = d_closed;
// Controller internal variables
\textbf(int) door_timer = 0;
// Test suite variables
\textbf(int) _test_step = 0, _test_index = -1;

\textbf(init) { \textbf(do) :: \textbf(atomic) {
  \textbf(if) // Test suite: test case selection
  :: _test_index == -1 ->
    \textbf(if)
    :: _test_index = 0;
    :: _test_index = 1;
    \textbf(fi)
  :: \textbf(else) -> ;
  \textbf(fi)
  \textbf(if) // Test suite: variable value selection
  :: _test_index == 0 ->
    \textbf(if)
    :: _test_step == 0 -> ;
    :: \textbf(else) ->
      user_floor_button[0] = 1;
      user_floor_button[1] = 1;
    \textbf(fi)
  :: \textbf(else) ->
    \textbf(if)
    :: _test_step == 0 ->
      user_cabin_button[1] = 1;
    :: \textbf(else) ->
      user_floor_button[2] = 1;
    \textbf(fi)
  \textbf(fi)
  _test_step = \lparen()_test_step + 1\rparen() % 2;
    
  \textbf(d_step) { // Plant execution
    elevator_pos = elevator_pos + up - down;
    elevator_pos = \lparen()elevator_pos > 6 -> 6 :
      elevator_pos\rparen();
    elevator_pos = \lparen()elevator_pos < 0 -> 0 :
      elevator_pos\rparen();
    \textbf(int) f; // floor
    \textbf(for) \lparen()f : 0..2\rparen() {
      on_floor[f] = elevator_pos == 3 * f;
      door_state[f] = \lparen()open[f] ->
        \lparen()door_state[f] == d_closed || door_state[f]
        == d_closing -> d_opening : d_open\rparen() :
        \lparen()door_state[f] == d_open || door_state[f]
        == d_opening -> d_closing : d_closed\rparen()\rparen();
      door_closed[f] = door_state[f] == d_closed;
      door_open[f] = door_state[f] == d_open;
      button[f] = \lparen()on_floor[f] && door_open[f] -> 0 :
        \lparen()user_f_button[f] -> 1 : button[f]\rparen()\rparen();
      call[f] = \lparen()on_floop[f] && door_open[f] -> 0 :
        \lparen()user_cabin_button[f] -> 1 : call[f]\rparen()\rparen();
    }
  }
 
  \textbf(d_step) { // Controller execution
    // <60 lines of code are omitted>
  }
} \textbf(od) }

\textbf(ltl) ERT\_1\_1 { \textbf(X)\lparen()[]\lparen()!on_floor[0] && !on_floor[1] &&
  !on_floor[2] -> door_closed[0]\rparen()\rparen() }
\end{Verbatim}

\subsection{Evaluation and comparison with model checking}
\label{sec:comp_with_mc}

The proposed testing framework was evaluated and compared with model checking.
For elevator models, the bound $\ell_{\max}$ for test case length was set to $3n + 6$, as $3(n - 1)$ test elements are required for the elevator to reach the highest position, and the remaining 9 elements are required to open and close the doors---this realizes not only state coverage, but also subformula coverage.
For PnP models, we empirically determined that $\ell_{\max} = 11$ is sufficient for $2 \le n \le 4$, and $\ell_{\max} = 17$ is sufficient for $n = 5$.

All experiments were performed on the Intel Core i7-4510U CPU with the clock rate of 2~GHz.
Table~\ref{tab:times} presents the results of applying the proposed testing framework to both manual and generated elevator and PnP models for various $n$: execution times of test generation in NuSMV, test execution in SPIN, and the total time of applying the framework (the sum of the first two times) are shown.
Elevator models with $n \le 5$ are omitted since they do not represent interest due to fast termination of all compared approaches.
On the other hand, with the growth of $n$, test generation time starts differing significantly between manual and generated models.
In particular, test generation did not terminate within the time limit of 12 hours for generated elevator models, which are more complex according to Table~\ref{tab:complexity}, with $n \ge 14$.

All the requirements ERT$_m^i, 1 \le m \le 4, 1 \le i \le n$ and PRT$_m^j, 1 \le m \le 3, 1 \le j \le 2^n - 1$ for all $n$ (except unfinished framework runs) were found to be satisfied.
However, since the proposed framework is an inexact technique, model checking on test cases may miss violations of some properties---the numbers of such violations are also shown in the table.
The overall number of properties which can potentially be falsely reported to be true, for elevator models, is $4n$---this is the number of properties ERF$_m^i$.
For PnP models, this number is $3 \cdot 2^n - 4$, which is the number of properties PRF$_m^j$.

\begin{table*}
\centering
\caption{The performance of the framework and usual model checking on the Elevator and the PnP case studies}
\renewcommand{\arraystretch}{1.3}
\label{tab:times}
\begin{tabular}{rrrrrrrrrrrrrrrr}
\hline\myskip
& & \multirow{2}{*}{\hspace{-0.2cm}$n$} & Test suite & \hspace{-0.3cm} & \multicolumn{3}{c}{Framework time (s)} & \hspace{-0.3cm} & \multicolumn{4}{c}{Model checking time (s)} & \hspace{-0.3cm} & \multicolumn{2}{c}{Missed property violations}\\
& & & \multicolumn{1}{c}{size$^{\text{a}}$} & \hspace{-0.3cm} & \hspace{-0.2cm}GEN$^{\text{b}}$ & EXEC$^{\text{c}}$ & Total & \hspace{-0.3cm} & \hspace{-0.1cm}CTL$^{\text{d}}$ & LTL$^{\text{e}}$ & $\frac{1}{2}$BMC$^{\text{f}}$ & BMC$^{\text{g}}$ & \hspace{-0.3cm} & \hspace{-0.1cm}Framework & $\frac{1}{2}$BMC$^{\text{f}}$ \\
\cline{1-4}\cline{6-8}\cline{10-13}\cline{15-16}\myskip
\multirow{14}{*}{\rotatebox[origin=c]{90}{Manually prepared models}} & \multirow{10}{*}{\rotatebox[origin=c]{90}{Elevator}}
  &  \hspace{-0.2cm}6 &  23/240 & \hspace{-0.3cm} &   \hspace{-0.1cm}4 & 12 &  16 & \hspace{-0.3cm} &       \hspace{-0.1cm}10 &       80 &   7 &   45 & \hspace{-0.3cm} &  \hspace{-0.1cm}1 &  7 \\
& &  \hspace{-0.2cm}7 &  27/324 & \hspace{-0.3cm} &   \hspace{-0.1cm}6 & 15 &  21 & \hspace{-0.3cm} &       \hspace{-0.1cm}39 &      485 &  13 &   95 & \hspace{-0.3cm} &  \hspace{-0.1cm}1 &  8 \\
& &  \hspace{-0.2cm}8 &  31/420 & \hspace{-0.3cm} &   \hspace{-0.1cm}9 & 18 &  27 & \hspace{-0.3cm} &       \hspace{-0.1cm}26 &      197 &  20 &  191 & \hspace{-0.3cm} &  \hspace{-0.1cm}1 &  9 \\
& &  \hspace{-0.2cm}9 &  35/528 & \hspace{-0.3cm} &  \hspace{-0.1cm}13 & 22 &  35 & \hspace{-0.3cm} &       \hspace{-0.1cm}24 &      166 &  40 &  339 & \hspace{-0.3cm} &  \hspace{-0.1cm}2 & 10 \\
& & \hspace{-0.2cm}10 &  39/648 & \hspace{-0.3cm} &  \hspace{-0.1cm}18 & 27 &  45 & \hspace{-0.3cm} &    \hspace{-0.1cm}17509 &     3525 &  62 &  604 & \hspace{-0.3cm} &  \hspace{-0.1cm}1 & 11 \\
& & \hspace{-0.2cm}11 &  43/780 & \hspace{-0.3cm} &  \hspace{-0.1cm}24 & 35 &  59 & \hspace{-0.3cm} &      \hspace{-0.1cm}731 & TL$^{\text{h}}$ & 127 & 1035 & \hspace{-0.3cm} &  \hspace{-0.1cm}0 & 12 \\
& & \hspace{-0.2cm}12 &  47/924 & \hspace{-0.3cm} &  \hspace{-0.1cm}34 & 42 &  77 & \hspace{-0.3cm} &       \hspace{-0.1cm}71 &     3512 & 186 & 2176 & \hspace{-0.3cm} &  \hspace{-0.1cm}1 & 13 \\
& & \hspace{-0.2cm}13 & 51/1080 & \hspace{-0.3cm} &  \hspace{-0.1cm}42 & 44 &  87 & \hspace{-0.3cm} &     \hspace{-0.1cm}1355 &    19897 & 324 & 3022 & \hspace{-0.3cm} &  \hspace{-0.1cm}3 & 14 \\
& & \hspace{-0.2cm}14 & 56/1249 & \hspace{-0.3cm} &  \hspace{-0.1cm}71 & 64 & 136 & \hspace{-0.3cm} &     \hspace{-0.1cm}8130 & TL$^{\text{h}}$ & 389 & 3784 & \hspace{-0.3cm} &  \hspace{-0.1cm}2 & 15 \\
& & \hspace{-0.2cm}15 & 59/1428 & \hspace{-0.3cm} & \hspace{-0.1cm}106 & 77 & 183 & \hspace{-0.3cm} & \hspace{-0.1cm}TL$^{\text{h}}$ & TL$^{\text{h}}$ & 557 & 6508 & \hspace{-0.3cm} &  \hspace{-0.1cm}2 & 16 \\
\cline{2-4}\cline{6-8}\cline{10-13}\cline{15-16}\myskip
& \multirow{4}{*}{\rotatebox[origin=c]{90}{PnP}}
  & \hspace{-0.2cm}2 &  7/25     & \hspace{-0.3cm} & \hspace{-0.1cm} 1    & 5     &    6  & \hspace{-0.3cm} &    \hspace{-0.1cm}2          &        9       &    1    &         4   & \hspace{-0.3cm} &  \hspace{-0.1cm}4    &  6   \\
& & \hspace{-0.2cm}3 &  11/40    & \hspace{-0.3cm} & \hspace{-0.1cm} 2    & 15    &   17  & \hspace{-0.3cm} &    \hspace{-0.1cm}268        &     4591       &    8    &        55   & \hspace{-0.3cm} &  \hspace{-0.1cm}8    & 14   \\
& & \hspace{-0.2cm}4 &  17/82    & \hspace{-0.3cm} & \hspace{-0.1cm} 6    & 43    &   49  & \hspace{-0.3cm} &    \hspace{-0.1cm}TL$^{\text{h}}$ &    TL$^{\text{h}}$ &  111   &       853   & \hspace{-0.3cm} &  \hspace{-0.1cm}16   & 15   \\
& & \hspace{-0.2cm}5 &  26/215 & \hspace{-0.3cm} & \hspace{-0.1cm}150 & 236 & 386 & \hspace{-0.3cm} &    \hspace{-0.1cm}TL$^{\text{h}}$ &     TL$^{\text{h}}$ &  3383 &  TL$^{\text{h}}$ & \hspace{-0.3cm} &  \hspace{-0.1cm}32 & 24 \\
\cline{1-4}\cline{6-8}\cline{10-13}\cline{15-16}\myskip
\multirow{14}{*}{\rotatebox[origin=c]{90}{Generated models}} & \multirow{10}{*}{\rotatebox[origin=c]{90}{Elevator}}
  &  \hspace{-0.2cm}6 &  24/242 & \hspace{-0.3cm} &       \hspace{-0.2cm}12 & 10 &    23 & \hspace{-0.3cm} &       \hspace{-0.1cm}14 &      101 &       36 &      239 & \hspace{-0.3cm} &   \hspace{-0.1cm}2 &  7 \\
& &  \hspace{-0.2cm}7 &  33/331 & \hspace{-0.3cm} &       \hspace{-0.1cm}29 & 13 &    42 & \hspace{-0.3cm} &       \hspace{-0.1cm}39 &      548 &      125 &      780 & \hspace{-0.3cm} &   \hspace{-0.1cm}0 &  8 \\
& &  \hspace{-0.2cm}8 &  34/424 & \hspace{-0.3cm} &       \hspace{-0.1cm}73 & 16 &    89 & \hspace{-0.3cm} &       \hspace{-0.1cm}99 &     1851 &      446 &     2806 & \hspace{-0.3cm} &   \hspace{-0.1cm}0 &  9 \\
& &  \hspace{-0.2cm}9 &  39/533 & \hspace{-0.3cm} &      \hspace{-0.1cm}220 & 19 &   239 & \hspace{-0.3cm} &      \hspace{-0.1cm}697 &     4533 &     1858 &     9794 & \hspace{-0.3cm} &   \hspace{-0.1cm}0 & 10 \\
& & \hspace{-0.2cm}10 &  47/657 & \hspace{-0.3cm} &      \hspace{-0.1cm}732 & 24 &   756 & \hspace{-0.3cm} &     \hspace{-0.1cm}1818 &    34516 &     6413 &    34193 & \hspace{-0.3cm} &   \hspace{-0.1cm}1 & 11 \\
& & \hspace{-0.2cm}11 &  47/785 & \hspace{-0.3cm} &     \hspace{-0.1cm}3212 & 28 &  3240 & \hspace{-0.3cm} &     \hspace{-0.1cm}2618 &    22807 &    26221 & TL$^{\text{h}}$ & \hspace{-0.3cm} &   \hspace{-0.1cm}2 & 12 \\
& & \hspace{-0.2cm}12 &  50/928 & \hspace{-0.3cm} &     \hspace{-0.1cm}7348 & 34 &  7381 & \hspace{-0.3cm} &    \hspace{-0.1cm}27258 & TL$^{\text{h}}$ & TL$^{\text{h}}$ & TL$^{\text{h}}$ & \hspace{-0.3cm} &   \hspace{-0.1cm}1 & 13 \\
& & \hspace{-0.2cm}13 & 54/1084 & \hspace{-0.3cm} &    \hspace{-0.1cm}29874 & 43 & 29917 & \hspace{-0.3cm} &     \hspace{-0.1cm}8543 & TL$^{\text{h}}$ & TL$^{\text{h}}$ & TL$^{\text{h}}$ & \hspace{-0.3cm} &   \hspace{-0.1cm}0 & 14 \\
& & \hspace{-0.2cm}14 &     --- & \hspace{-0.3cm} & \hspace{-0.1cm}TL$^{\text{h}}$ & ---&TL$^{\text{h}}$& \hspace{-0.3cm} & \hspace{-0.1cm}TL$^{\text{h}}$ & TL$^{\text{h}}$ & TL$^{\text{h}}$ & TL$^{\text{h}}$ & \hspace{-0.3cm} & \hspace{-0.1cm}--- & ---\\
& & \hspace{-0.2cm}15 &     --- & \hspace{-0.3cm} & \hspace{-0.1cm}TL$^{\text{h}}$ & ---&TL$^{\text{h}}$& \hspace{-0.3cm} & \hspace{-0.1cm}TL$^{\text{h}}$ & TL$^{\text{h}}$ & TL$^{\text{h}}$ & TL$^{\text{h}}$ & \hspace{-0.3cm} & \hspace{-0.1cm}--- & ---\\
\cline{2-4}\cline{6-8}\cline{10-13}\cline{15-16}\myskip
& \multirow{4}{*}{\rotatebox[origin=c]{90}{PnP}}
  & \hspace{-0.2cm}2 &  9/31 & \hspace{-0.3cm} & \hspace{-0.1cm}1 & 5 & 6 & \hspace{-0.3cm} &    \hspace{-0.1cm}1 &     3 &  1 &  7 & \hspace{-0.3cm} &  \hspace{-0.1cm}4 & 6 \\
& & \hspace{-0.2cm}3 &  13/45 & \hspace{-0.3cm} & \hspace{-0.1cm}3 & 12 & 15 & \hspace{-0.3cm} &    \hspace{-0.1cm}73 &     1588 &  13 &  79 & \hspace{-0.3cm} &  \hspace{-0.1cm}7 & 14 \\
& & \hspace{-0.2cm}4 &  32/104 & \hspace{-0.3cm} & \hspace{-0.1cm}31 & 38 & 69 & \hspace{-0.3cm} &    \hspace{-0.1cm}TL$^{\text{h}}$ &     TL$^{\text{h}}$ &  704 &  4729 & \hspace{-0.3cm} &  \hspace{-0.1cm}15 & 14 \\
& & \hspace{-0.2cm}5 &  65/164 & \hspace{-0.3cm} & \hspace{-0.1cm}3432 & 150 & 3582 & \hspace{-0.3cm} &    \hspace{-0.1cm}TL$^{\text{h}}$ &     TL$^{\text{h}}$ &  TL$^{\text{h}}$ &  TL$^{\text{h}}$ & \hspace{-0.3cm} &  \hspace{-0.1cm}31 & --- \\
\hline
\multicolumn{16}{l}{\scriptsize $^{\text{a}}$Given in the format $<$number of test cases$>$/$<$total number of elements$>$.
$^{\text{b}}$Test generation time. $^{\text{c}}$Test execution time.}\\
\multicolumn{16}{l}{\scriptsize $^{\text{d}}$BDD-based CTL model checking. $^{\text{e}}$BDD-based LTL model checking. $^{\text{f}}$LTL BMC with $k = \lfloor k_{\text{opt}} / 2 \rfloor$. $^{\text{g}}$LTL BMC with $k = k_{\text{opt}}$.}\\
\multicolumn{16}{l}{\scriptsize $^{\text{h}}$Time limit (12 hours) was exceeded.}
\end{tabular}
\end{table*}


Conventional model checking was also performed.
Explicit-state model checking became impractically long for $n \ge 5$, so its results are not reported.
Since the requirements are available in both LTL and CTL, the cases of both LTL and CTL symbolic BDD-based model checking were examined.
Then, BMC is only possible for LTL properties, and represents particular interest for comparison since this is a technique which preserves the verified system but verifies it less thoroughly than BDD-based model checking.
If the maximum possible bound for counterexamples $k_{\text{opt}}$ is known, then BMC becomes precise.
Determining $k_{\text{opt}}$ may be difficult for arbitrary LTL properties and formal models, but for the requirements of types ERF$_m^i$ and PRF$_m^j$ this can be done relatively easy.
Corresponding calculations are omitted to save space, but the results are as follows:
\begin{enumerate}
\item for elevator models, $k_{\text{opt}} = 3n + 5$ and is realized on requirements of the type ERF$_4^i$;
\item for PnP models, $k_{\text{opt}} = 2^n + 12$ and is realized on requirements of the type PRF$_1^j$.
\end{enumerate}

In addition to the bound $k = k_{\text{opt}}$, a smaller one, $k = \lfloor k_{\text{opt}} / 2 \rfloor$ (the brackets denote rounding down), was also used to obtain lower BMC execution times which would be comparable with the ones of applying the framework.
On elevator models, BMC with such a bound can be shown to miss violations of exactly $n + 1$ temporal properties out of $4n$ false ones, where missed violations are limited to requirements of types ERF$_3^i$ and ERF$_4^i$.
The results of comparing the framework with various types of model checking are also given in Table~\ref{tab:times}.


Speaking of elevator models, for large $n$ our examples become challenging for exact model checking.
Execution times of BDD-based model checking are low in some cases, but are unstable, which is especially visible on manual models.
In contrast, BMC execution times grow predictably, but faster than the ones of applying the framework.
In almost all experiments, BMC with $k = \lfloor k_{\text{opt}} / 2 \rfloor$ is slower than the framework; what is more, it misses more violations of temporal properties compared to the framework.
We also note that although sometimes CTL model checking outperforms the framework, LTL is more common in the automation systems context, and only particular LTL requirements can be represented in CTL (this case study was designed to make it possible for all the used requirements).

On PnP models, BDD-based verification is clearly slower than the other approaches, although the number of experiments is less.
The execution time of the testing framework was again lower than the one of BMC with $k = \lfloor k_{\text{opt}} / 2 \rfloor$, although now it misses more property violations.

\section{Comparison with other approaches of integrating testing and model checking}
\label{sec:comparison}

Multiple approaches of test case generation were proposed in previous studies~\cite{fraser2009testing}.
The framework proposed in the present paper uses the most common and computationally simple ideas of these approaches, but is not a test case generation approach itself.
Yet, it includes a simple technique to cover subformulas of temporal specifications, whose use needs to be justified in comparison with the approach described in~\cite{tan2004specification}.
In this approach, the used coverage criterion is more elaborate than simple reachability of subformula values: intuitively, it requires the test cases to exclude different mutations of the LTL specification $f$.
To achieve such coverage, for each atomic proposition of $f$, a model checker is run on a modified version of $f$.
However, the goal of our testing framework is to overcome the computational difficulty of verifying $f$, which is impossible with such a procedure.
This justifies the use of our simpler technique, although its utility with respect to test suite reliability is clearly lower.

The work~\cite{buzhinsky2015formal} reduces the problem of testing to the one of model checking.
However,
in this method each test case is encoded as a list of input/output pairs: for a given sequence of inputs from the plant the output sequence required from the controller is fixed.
Specifying the output behavior precisely limits the capabilities of testing.
In contrast, temporal specifications used in the present paper rather forbid patterns of undesired input/output sequences.
Then, the work~\cite{buzhinsky2015formal} employs the formalism of net condition/event systems (NCES), the maintenance of tools for which has finished.

The contributions of~\cite{buzhinsky2015formal} are developed in~\cite{buzhinsky2017testing}, where NuSMV is used as a modeling formalism.
The general test case execution idea of~\cite{buzhinsky2017testing} is partially adopted in the present work, but the following differences are notable:
\begin{enumerate}
\item In~\cite{buzhinsky2017testing}, the nondeterminism of the closed-loop model is achieved by making the plant model nondeterministic, rather than extracting nondeterministic variables.
\item In~\cite{buzhinsky2017testing}, test cases are represented as finite-state acceptors which are obtained using MBT, instead of input sequences in the present work.
Consequently, (1)~test oracles are specified within test cases, not separately in temporal specifications, and (2)~test case generation is not fully automatic.
\item Although the work~\cite{buzhinsky2017testing} considers test case execution for closed-loop SUVs, it does not propose a formal method of quality assurance, but rather explores different use cases of test case execution by model checkers assuming that the test suite is given.
\end{enumerate}


To sum up, the proposed framework differs from the aforementioned approaches by binding two research directions together: test case generation and execution using model checkers.
This connection has not been done previously and includes the following novel elements: (1)~support of closed-loop SUVs, (2)~isolation of nondeterminism into separate variables, (3)~the selection of computationally simple coverage criteria for the controller model, the plant model and the temporal specification, (4)~the use of temporal requirements as test oracles, and (5)~looping test cases to maintain the semantics of the temporal requirements unchanged during test case execution.

One more approach is built on top of ideas different from the ones discussed above.
In~\cite{duarte2011integrating}, testing and model checking are combined in a loop of incremental model generation: the model of the system is generated based on behavior traces, then this model is verified, and obtained counterexamples are added to the test suite comprised of behavior traces.
Thus, this approach incrementally improves both the test suite and the model of the system.
The work~\cite{duarte2011integrating} differs from the present one by the employed methodology: the system is treated as a black box rather than a white box.
The benefit of such an approach is the lack of need of a formal model.

Finally, runtime verification~\cite{leucker2009brief} is a technique of checking properties during SUV execution by supplementing the SUV with correctness monitors which execute in parallel with the SUV.
Consequently, differently from model checking, conclusions of runtime verification must be drawn based on finite executions of the SUV.
On the other hand, runtime verification can be run on the actual SUV or, in the case of automation systems, on the model of SUV which is richer than the one prepared for model checking (such as a simulation model).
Properties to be checked in runtime verification are based on system specification and, in particular, can be equivalent to the ones representable in LTL.
However, due to the need to observe finite traces, checking some LTL properties (\emph{liveness} ones) becomes impossible.
The question of selecting simulations to be executed is similar to the one of test case selection.
To sum up, runtime verification is a testing approach which avoids consideration of formal models and thus is significantly different from the one proposed in this paper.




\section{Discussion and conclusions}
\label{sec:discussion}

In this paper, a framework for industrial automation systems has been proposed that methodologically lies between testing and formal verification.
Viewed as a method of testing, the framework incorporates an approach to generate test suites based on coverage with oracles formulated as temporal properties.
To generate and run such test suites, the SUV is examined at the formal level, which is not the case in conventional testing.
As shown in~\cite{buzhinsky2017testing}, such an examination has a number of potential benefits, including the lack of the need to wait until the SUV executes.
The entirely formal view also differentiates the framework from runtime verification.
Viewed as a method of formal verification, the framework binds previously proposed approaches of test case generation and execution using model checking and becomes a complement to existing techniques which reduce model checking complexity.
Like BMC, it suggests a reduced form of verification, searching for a trade-off between verification reliability and feasibility, but the approach to reduction is different.

However, achieving any sort of coverage cannot guarantee that the generated test cases would contain counterexamples to all temporal properties which are actually violated for the SUV.
It is rather advised to apply the framework together with BMC, runtime verification and various abstraction techniques.
To conclude, the framework represents a verification technique different from existing ones and hence can supplement them in ensuring reliability of automation systems.

Two ideas proposed in the framework deserve separate discussion.
First, symbolic and explicit-state model checkers have been used to achieve different purposes: while symbolic model checker NuSMV was needed to generate test cases, explicit-state model checker SPIN was used to execute them.
The shortcoming of this approach is that both NuSMV and Promela formal models are needed for the considered system.
They may be difficult to obtain since these languages require declarative and imperative model description, respectively.
On the other hand, such models can be obtained automatically, as demonstrated for the case of generated elevator models.
Another possible solution may involve using a verifier which supports both symbolic and explicit-state model checking, such as LTSmin~\cite{kant2015ltsmin}.

The second idea is the one of extracting the nondeterminism of the plant model into separate variables.
One can imagine a more straightforward solution: the plant is modeled as a nondeterministic state machine, test cases are generated as sequences of input and output values, and then these sequences are checked against the temporal requirements.
The reason of not choosing this path is efficiency:
\begin{enumerate}
\item BMC would become unusable for test case generation as producing finite input/output sequences which cannot be extended to infinite ones (looping would create incorrect plant/controller interactions);
\item the applicability of the cone of influence reduction of NuSMV would be greatly lowered as much more variables need to be included into test cases;
\item this verbosity of test suites would also slow down model compilation in SPIN, which is done prior to verification.
\end{enumerate}


The performed study has several limitations which can be addressed in future work.
First, the issue of determining a suitable test case length during test case generation has not been considered in the general case.
A possible solution is to increase it incrementally until system coverage stops growing.
Then, applying the framework to check liveness properties, which require infinite behavior traces to show their violation, may be problematic as the transformation of finite test cases to infinite ones is by now somewhat arbitrary: test cases are looped infinitely from the beginning.
Future work may also involve improving the developed tool by integrating it with approaches of automatic formal model generation, such as~\cite{darvas2017plc, gourcuff2006efficient, drozdov2017formal}, and user-friendly model checking, including counterexample explanation techniques~\cite{beer2012explaining}.
The latter would help the user to understand what happens in the SUV when a violated test case is executed, and whether this violation is connected with modeling or reveals an actual fault in the SUV.

\bibliographystyle{spmpsci}

\bibliography{paper}\vspace{0cm}
\end{document}